\documentstyle[emulateapj,epsf]{article}

\def\ltsima{$\; \buildrel < \over \sim \;$}
\def\simlt{\lower.5ex\hbox{\ltsima}}
\def\gtsima{$\; \buildrel > \over \sim \;$}
\def\simgt{\lower.5ex\hbox{\gtsima}}
 
\received{}
\accepted{}
\journalid{}{}
\articleid{}{}
\slugcomment{accepted for publication in ApJ-Letters}
\lefthead{}
\righthead{}

\begin{document}
   
\title{A BeppoSAX observation of the merging cluster Abell 3266}
\author{Sabrina De Grandi\altaffilmark{1} \& Silvano Molendi\altaffilmark{2}
}

\altaffiltext{1}{Osservatorio Astronomico di Brera, via Bianchi 46,
I-23807 Merate (LC), Italy}

\altaffiltext{2}{Istituto di Fisica Cosmica, CNR, via Bassini 15,
I-20133 Milano, Italy}

\begin{abstract}
We present results from a BeppoSAX observation of the rich cluster
Abell 3266.  The broad band spectrum (2-50 keV) of the cluster, when
fitted with an optically thin thermal emission model, yields a
temperature of 8.1$\pm$0.2 keV and a metal abundance of 0.17$\pm$0.02
in solar units, and with no evidence of a hard X-ray excess in the PDS
spectrum.  By performing a spatially resolved spectral analysis we
find that the projected temperature drops with increasing radius,
going from $\sim$ 10 keV at the cluster core to $\sim$ 5 keV at about
1.5 Mpc.  Our BeppoSAX temperature profile is in good agreement with
the ASCA temperature profile of Markevitch et al. (1998).  From our
two-dimensional temperature map we find that the gradient is observed
in all azimuthal directions.  The temperature gradient may have been
caused by a recent merger event also responsible for a velocity
dispersion gradient measured in the optical band.  The projected metal
abundance profile and two-dimensional map are both consistent with
being constant.

\end{abstract}


\section {Introduction}

Abell 3266 (hereafter A3266), also known as Sersic 40/6, is a rich,
nearby (z$=$ 0.055, Teague et al. 1990), cluster of galaxies.  It has
been extensively studied at both optical and X-ray wavelengths.  In
the optical band various authors have studied the dynamics of this
cluster by analyzing the velocity dispersion of a large number of
galaxies (e.g. Teague et al. 1990, 152 galaxies; Quintana, Ramirez
\& Way 1996, hereafter QRW, 387 galaxies). QRW found evidence of a 
decrease of the velocity dispersion with increasing distance from the
cluster core.  Similar results were also reported by Girardi et
al. (1997).  The velocity dispersion radial gradient and the presence
of a distorted central dumb-bell galaxy have been interpreted by QRW as
evidence of a recent merger along the NE-SW direction. According to
the above authors the two subclusters started colliding about 4 Gyr
ago, with the central cores coming together in the last 1-2 Gyr.

X-ray observations with the Einstein HRI (Mohr, Fabricant \& Geller
1993), and the ROSAT PSPC (Mohr, Mathiesen \& Evrard 1999), have
confirmed that A3266 is far from being a relaxed cluster.  The
isophotes on the few hundred kpc scale are elongated in the NE-SW
direction, while on the few Mpc scales the elongation shifts to the
E-W direction. The azimuthally averaged surface brightness profile
(see figure 9 of Mohr, Mathiesen \& Evrard 1999), is characterized by
a relatively large core radius of $\sim$ 500 kpc and is not well
fitted by a $\beta$ model, confirming the non-relaxed status of this
cluster.  Peres et al. (1998), by applying the deprojection technique
(Fabian et al. 1980), found no evidence of a cooling flow in the core
of A3266.  David et al. (1993), using Einstein MPC data, report a
global temperature of 6.2$^{+0.5}_{-0.4}$ keV for A3266.  Markevitch
et al. (1998), from the analysis of ASCA data, found evidence of a
strong temperature gradient in A3266.  The projected temperature was
found to decrease from $\sim$ 10 keV to $\sim$ 5 keV when going from
the cluster core out to $\sim$ 1.5 Mpc.  Temperature maps of A3266
(Markevitch et al. 1998; Henricksen, Donnelly \& David 1999) indicate
an asymmetric temperature pattern, which could be associated with the
ongoing merger.  Irwin, Bregman \& Evrard (1999), who have used ROSAT
PSPC data to search for temperature gradients in a sample of galaxy
clusters including A3266, in contrast with Markevitch et al. (1998),
did not find any evidence of a temperature gradient in A3266.
Mushotzky (1984), using HEAO1 A2 data, found a value of 0.4$\pm$0.2,
solar units, for the Fe abundance of A3266.

In this Letter we report a recent BeppoSAX observation of A3266.  We
use our data to perform an independent measurement of the temperature
profile and two-dimensional map of A3266.  We also present the first
abundance profile and map of A3266 and the first measurement of the
hard (15-50 keV) X-ray spectrum of A3266.  The outline of the Letter
is as follows.  In section 2 we give some information on the BeppoSAX
observation of A3266 and on the data preparation.  In section 3 we
present the analysis of the broad band spectrum (2-50 keV) of A3266.
In section 4 we present spatially resolved measurements of the
temperature and metal abundance.  In section 5 we discuss our results
and compare them to previous findings.  
Throughout this Letter we assume H$_{o}$=50 km s$^{-1}$Mpc$^{-1}$ and
q$_{o}$=0.5.
 
\section {Observation and Data Preparation}
The cluster A3266 was observed by the BeppoSAX satellite (Boella et
al. 1997a) between the 24$^{th}$ and the 26$^{st}$ of March 1998.  We
will discuss here data from two of the instruments onboard BeppoSAX:
the MECS and the PDS.  The MECS (Boella et al. 1997b)
is presently composed of two units
working in the 1--10 keV energy range. At 6~keV, the energy resolution
is $\sim$8\% and the angular resolution is $\sim$0.7$^{\prime}$
(FWHM). The PDS instrument (Frontera et al. 1997), is a passively
collimated detector (about 1.5$\times$1.5 degrees f.o.v.), working in
the 13--200 keV energy range.  Standard reduction procedures and
screening criteria have been adopted to produce linearized and
equalized event files.  Both MECS and PDS data preparation and
linearization was performed using the {\sc Saxdas} package under {\sc
Ftools} environment.  The effective exposure time of the observation
was 7.6$\times$10$^4$ s (MECS) and 3.2$\times$10$^4$ s (PDS).  The
observed countrate for A3266 was 0.488$\pm$0.003 cts/s for the 2 MECS
units and 0.23$\pm$0.03 cts/s for the PDS instrument.

All MECS spectra discussed in this Letter have been background
subtracted using spectra extracted from blank sky event files in the same
region of the detector as the source.
All spectral fits have been performed using XSPEC Ver. 10.00.  Quoted
confidence intervals are 68$\%$ for 1 interesting parameter
(i.e. $\Delta \chi^2 =1$), unless otherwise stated.

\section{Broad Band Spectroscopy} 

We have extracted a MECS spectrum, in the 2-10 keV band, from a
circular region of 14$^{\prime}$ radius (1.2 Mpc), centered on the 
emission peak. From the ROSAT PSPC radial
profile, we estimate that about 89\% of the total cluster emission
falls within this radius.  
The PDS ($13-50$ keV)
background-subtracted spectrum has been produced by subtraction of the
``off-'' from the ``on-source'' spectrum.
The spectra from the two instruments have been fitted simultaneously
with an optically thin thermal emission model (MEKAL code in the XSPEC 
package), absorbed by a galactic line of sight equivalent hydrogen 
column density, $N_H$, of 1.6$\times 10^{20}$ cm$^{-2}$
(Dickey \& Lockman 1990).    
A numerical relative normalization factor among the two
instruments has been added to account for: 
a) the fact that the MECS spectrum includes emission out to 1.2 Mpc from 
the X-ray peak, while the PDS field of view (1.3 degrees FWHM) covers the 
entire emission from the cluster;
b) the slight mismatch in the absolute flux calibration
of the MECS and PDS response matrices employed in this Letter 
(September 1997 release); 
c) the vignetting in the PDS instrument, 
(the MECS vignetting is included in the response matrix thanks to the 
{\sc Effarea} program described in the following section).
The estimated normalization factor is 0.9. In the fitting procedure we
allow this factor to vary within 15$\%$ from the above value to account
for the uncertainty in this parameter.  The MEKAL model yields an
acceptable fit to the data, $\chi^2 =$ 191.3 for 176 d.o.f. The best
fitting values for the temperature and the metal abundance are
8.1$\pm$0.2 keV and 0.17$\pm$0.02 respectively, where the latter value
is expressed in solar units.  In figure 1 we show the MECS and PDS
spectra of A3266 together with the best fitting model.
The PDS data shows no evidence of a hard X-ray excess.

\section{Spatially Resolved Spectral Analysis} 

When performing spatially resolved spectral analysis of galaxy
clusters one must take into account the distortions introduced by the
energy dependent PSF.  In the case of the MECS instrument onboard
BeppoSAX, the PSF is found to vary only weakly with energy (D'Acri, De
Grandi \& Molendi 1998), and therefore the spectral distortions are
expected to be small. Nonetheless they have been taken into account
using the {\sc Effarea} program publicly available within the latest
{\sc Saxdas} release.  As explained in Molendi et al. (1999),
hereafter M99, the {\sc Effarea} program convolves the ROSAT PSPC
surface brightness with an analytical model of the MECS PSF (see
D'Acri, De Grandi \& Molendi 1998, for a more extensive description).
The {\sc Effarea} program also includes corrections for the energy
dependent telescope vignetting, which are not discussed in D'Acri et
al. (1998). The {\sc Effarea} program produces effective area files,
which can be used to fit spectra accumulated from annuli or from
sectors of annuli.

\subsection{Radial Profiles}

We have accumulated spectra from 7 annular regions centered on the
X-ray emission peak, with inner and outer radii of
0$^{\prime}$-2$^{\prime}$, 2$^{\prime}$-4$^{\prime}$,
4$^{\prime}$-6$^{\prime}$, 6$^{\prime}$-8$^{\prime}$,
8$^{\prime}$-12$^{\prime}$, 12$^{\prime}$-16$^{\prime}$ and
16$^{\prime}$-20$^{\prime}$.  
A correction for the
absorption caused by the strongback supporting the detector window has
been applied for the 8$^{\prime}$-12$^{\prime}$ annulus, where the
annular part of the strongback is contained. For the
4$^{\prime}$-6$^{\prime}$, 12$^{\prime}$-16$^{\prime}$ and the
16$^{\prime}$-20$^{\prime}$ annuli, where the strongback covers only a
small fraction of the available area, we have chosen to exclude the
regions shadowed by the strongback.  For the 5 innermost annuli the
energy range considered for spectral fitting was 2-10 keV, while for
the 2 outermost annuli, the fit was restricted to the 2-8 keV energy
range.  We have used a softer energy range for the outer annuli to
limit spectral distortions which could be caused by an incorrect
background subtraction.  The MECS instrumental background has a very
hard spectrum that, in the outer regions, accounts for about 60$\%$
of the total intensity in the 8-10 keV band, and that can vary up to
10$\%$ from one observation to another.
 
We have fitted each spectrum with a MEKAL model absorbed by the
galactic $N_H$, of
1.6$\times 10^{20}$ cm$^{-2}$.  In figure 2 we show the temperature
and abundance profiles obtained from the spectral fits.  By fitting
the temperature and abundance profiles with a constant we derive the
following average values: $8.7\pm$0.3 keV and 0.21$\pm$0.03, solar
units.  A constant does not provide an acceptable fit to the
temperature profile.  Using the $\chi^2$ statistics we find: $\chi^2
=$ 17.7 for 6 d.o.f., corresponding to a probability of 0.007 for the
observed distribution to be drawn from a constant parent distribution.
A linear profile of the type, kT = a $ + $ b~r, where kT is in keV and
r in arcminutes, provides a much better fit, $\chi^2 =$ 0.75 for 5
d.o.f. The best fitting values for the parameters are a$ = 10.48 \pm
0.52$ keV, b$= - ~ 0.307 \pm 0.075$ keV~arcmin$^{-1}$.  A constant
provides an acceptable
fit to the abundance profile, $\chi^2 =$ 4.4 for 6
d.o.f. (Prob.$=$0.6).  
   
As in M99, we have used the Fe K$_{\alpha}$ line as an independent
estimator of the ICM temperature.  Briefly we recall that the centroid
of the observed Fe K$_{\alpha}$ line depends upon the relative
contributions of the He-like Fe line at 6.7 keV, and the H-like Fe
line at 7.0 keV. Since the relative strength of these two lines is a
function of the gas temperature, the centroid of the observed line is
also a function of the gas temperature.  Moreover, the position of the
centroid of the Fe K$_{\alpha}$ line is essentially unaffected by the
spectral distortion introduced by the energy dependent PSF and depends
only weakly on the the adopted continuum model.  Thus it can be used
to derive an independent and robust estimate of the temperature
profile.  Considering the limited number of counts available in the
line we have performed the analysis on 2 annuli with bounding radii,
0$^{\prime}$-8$^{\prime}$ and 8$^{\prime}$-16$^{\prime}$.  We have
fitted each spectrum with a bremsstrahlung model plus a line, both at
a redshift of z=0.055 (ZBREMSS and ZGAUSS models in XSPEC), absorbed
by the galactic $N_H$.  A systematic negative shift of 40 eV has been
included in the centroid energy to account for a slight
misscalibration of the energy pulseheight-channel relationship near
the Fe line. To
convert the energy centroid into a temperature we have derived an
energy centroid vs. temperature relationship.  This has been done by
simulating thermal spectra, using the MEKAL model and the MECS
response matrix, and fitting them with the same model, which has been
used to fit the real data.
In figure 2 we have overlaid the temperatures derived from the
centroid analysis on those previously obtained through the thermal
continuum fitting. The two measurements of the temperature profile are
in agreement with each other.  Unfortunately, the modest statistics
available in the line does not allow us to say much more than that.

\subsection{Maps}

We have divided A3266 into 4 sectors: NW, SW, SE and NE.  Each sector
has been divided into 3 annuli with bounding radii,
2$^{\prime}$-4$^{\prime}$, 4$^{\prime}$-8$^{\prime}$ and
8$^{\prime}$-16$^{\prime}$.  The orientation of the sectors has been
chosen so that the North-South division roughly coincides with the
apparent major axis of the X-ray isophotes.  In figure 3 we show the
MECS image with the sectors overlaid.  A correction for the absorption
caused by the strongback supporting the detector window has been
applied for the sectors belonging to the 8$^{\prime}$-16$^{\prime}$
annulus.  For the sectors in the 2$^{\prime}$-4$^{\prime}$ and
4$^{\prime}$-8$^{\prime}$ annuli, we used the 2-10 keV energy range
for spectral fitting, while for the 8$^{\prime}$-16$^{\prime}$ annulus
we adopted the 2-8 keV range.  We have fitted each spectrum with a
MEKAL model absorbed by the galactic $N_H$.

In figure 4 we show the temperature profiles obtained from the
spectral fits for each of the 4 sectors.  Note that in all the
profiles we have included the temperature measure obtained for the
central circular region with radius 2$^{\prime}$.  Fitting each radial
profile with a constant temperature we derive the following average
sector temperatures: 8.8$\pm$0.5 keV for the NW sector, 9.6$\pm$0.5
keV for the SW sector, 8.1$\pm$0.5 keV for the SE sector and
8.2$\pm$0.4 keV for the NE sector.  For all sectors we find a
statistically significant temperature decrease with increasing
radius. From the $\chi^2$ statistics we find $\chi^2 =21.4$ for 3
d.o.f. (Prob.$= 9\times 10^{-5}$) for the NW sector, $\chi^2 =9.2$ for
3 d.o.f. (Prob.$= 2.6\times 10^{-2}$) for the SW sector, $\chi^2
=24.0$ for 3 d.o.f. (Prob.$= 2.5\times 10^{-5}$) for the SE sector and
$\chi^2 =10.5$ for 3 d.o.f. (Prob.$= 1.5\times 10^{-2}$) for the NE
sector.  In the SE and NE sectors the temperature decreases
continuously as the distance from the cluster center increases. In the
NW and SW sectors the temperature first increases, reaching a maximum
in either the second (NW sector) or third (SW sector) annulus, and
then decreases.  Interestingly, a fit to the temperatures of the 4
sectors in the third annulus (bounding radii 4$^\prime$-8$^\prime$)
with a constant, yields $\chi^2=8.45$ for 3 d.o.f., with an associated
probability for the temperature to be constant of 0.03, indicating
that an azimuthal temperature gradient may be present near the core of
the cluster.  More specifically the eastern side of the cluster
appears to be somewhat cooler than the western side. 
From the analysis of the abundance map we find that all sector
averaged abundances are consistent with the average
abundance for A3266 derived in the previous subsection 
The $\chi^2$ values derived from the fits indicate that all abundance
profiles are consistent with being constant.

\section{Discussion}

Previous measurements of the temperature structure of A3266 have been
performed by Markevitch et al. (1998), using ASCA data, and by Irwin,
Bregman \& Evrard (1999), using ROSAT PSPC data.
Markevitch et al. (1998) find a decreasing radial temperature profile.
In figure 2 we have overlaid the temperature profile obtained by
Markevitch et al. (1998) using ASCA data, to our own BeppoSAX profile.
The agreement between the two independent measurements is clearly very
good.  A linear profile of the type, kT = a $ + $ b~r, where kT is in
keV and r in arcminutes, which provides an acceptable fit to the ASCA
profile ($\chi^2 =5\times 10^{-5}$ for 1 d.o.f.)  yields best fitting
values: a $= 10.7 \pm 0.8$ keV, b$= - 0.39 \pm 0.11$
keV~arcmin$^{-1}$.  These values are in good agreement with those
derived from the BeppoSAX data.
Recently Irwin, Bregman \& Evrard (1999) have used ROSAT PSPC hardness
ratios to measure temperature gradients for a sample of nearby galaxy
clusters, which includes A3266. In their analysis they find evidence
of a radial decrease in one of the two hardness ratios sensitive to
temperature variations. The authors do not attribute this variation to
a temperature decrement, because a similar variation is also seen in
an another hardness ratio, which is not sensitive to temperature
gradients.

Optical studies by various authors (e.g., QRW, Teague et al. 1990),
have shown that A3266 is characterized by a large velocity dispersion,
$\sim$ 1000 km s$^{-1}$.  Moreover both QRW and Girardi et al. (1997)
find evidence of a decrease of the velocity dispersion with increasing
distance from the cluster core.  QRW measure a velocity dispersion of
$\sim$ 1600 km s$^{-1}$ within 200 kpc from the core of the cluster
and a velocity dispersion of $\sim$ 1000 km s$^{-1}$ at a radial
distance of 18$^\prime$ (1.5 Mpc).  Thus, it would
seems that both the hot X-ray emitting gas and the galaxies visible at
optical wavelengths are characterized by a decrease in their specific
kinetic energy with increasing radius.  From the velocity dispersion
profile produced by QRW and our own temperature profile, we have
computed the radial profile of the so-called $\beta_{\rm spec}$
parameter (Sarazin 1988), which is defined as: $\rm {\beta_{\rm spec}
\equiv \mu m_p\sigma_r^2/kT}$, where $\mu$ is the mean molecular weight in amu,
m$_{\rm p}$ is the proton mass and $\sigma_{\rm r}$ is the velocity
dispersion.  The derived $\beta_{\rm spec}$ profile is consistent with
being flat, with an average value of $\beta_{\rm spec} = 1.5 \pm 0.2$,
indicating that, although the specific kinetic energy of the galaxies
is greater than that of the hot gas, the rate at which it decreases,
with increasing radius, is the same for the galaxies and the hot gas.
QRW have proposed that the velocity dispersion gradient and the
presence of a distorted central dumb-bell galaxy may have resulted
from a recent merger between two clusters.  Evidence of a merger event
can also be found from the X-ray data.  The ROSAT PSPC image (Mohr,
Mathiesen \& Evrard 1999), shows isophotes elongated in the NE-SW
direction on the few hundred kpc scale, while on the few Mpc scales
the elongation shifts to the E-W direction.  In the picture proposed
by QRW the two clusters started colliding about 4 Gyr ago, with the
central cores coming together in the last 1-2 Gyr.  Incidentally a fit
with a $\beta$-model to the PSPC radial profile (see Mohr, Mathiesen
\& Evrard 1999) yields a large value for the core radius, $r_c=0.5$
Mpc, as might expected in the case of recent merger in the core of
A3266.  The radial temperature gradient found by ASCA and BeppoSAX
lends further strength to the merging scenario proposed by QRW.

The map we present in figure 3 shows that the radial temperature
gradient is present in all sectors.  We also find an indication of an
azimuthal temperature gradient occurring in the annulus with bounding
radii 4$^\prime$-8$^\prime$ (0.35 Mpc - 0.7 Mpc); the data suggests
that the eastern side of the cluster may be somewhat cooler than the
western side.  Very recently Henriksen, Donnelly \& Davis (1999), from
the analysis of an ASCA observation of A3266, find evidence of a
temperature gradient in the SW$-$NE direction indicative of an ongoing
merger. The azimuthal temperature gradient found in our data
corroborates the ASCA result.

The average metal abundance we find from the MECS data, $0.21\pm
0.03$, solar units, is in agreement with the value 0.4$\pm$0.2 derived
by Mushotzky (1984), using HEAO1 A2 data, and with the average
metallicity, $0.21\pm 0.05$, derived by Allen \& Fabian (1998) for a
sample of non-cooling flow cluster.  The radial abundance profile (see
the bottom panel of figure 2), does not show any strong evidence of a
decrease in the abundance with increasing radius. Thus, A3266 would
seem to conform to the general rule that non-cooling flow cluster do
not present metallicity gradients.


\acknowledgments
We acknowledge support from the BeppoSAX Science Data Center.


\clearpage


{\begin{figure}
\epsfxsize=\textwidth
\epsffile{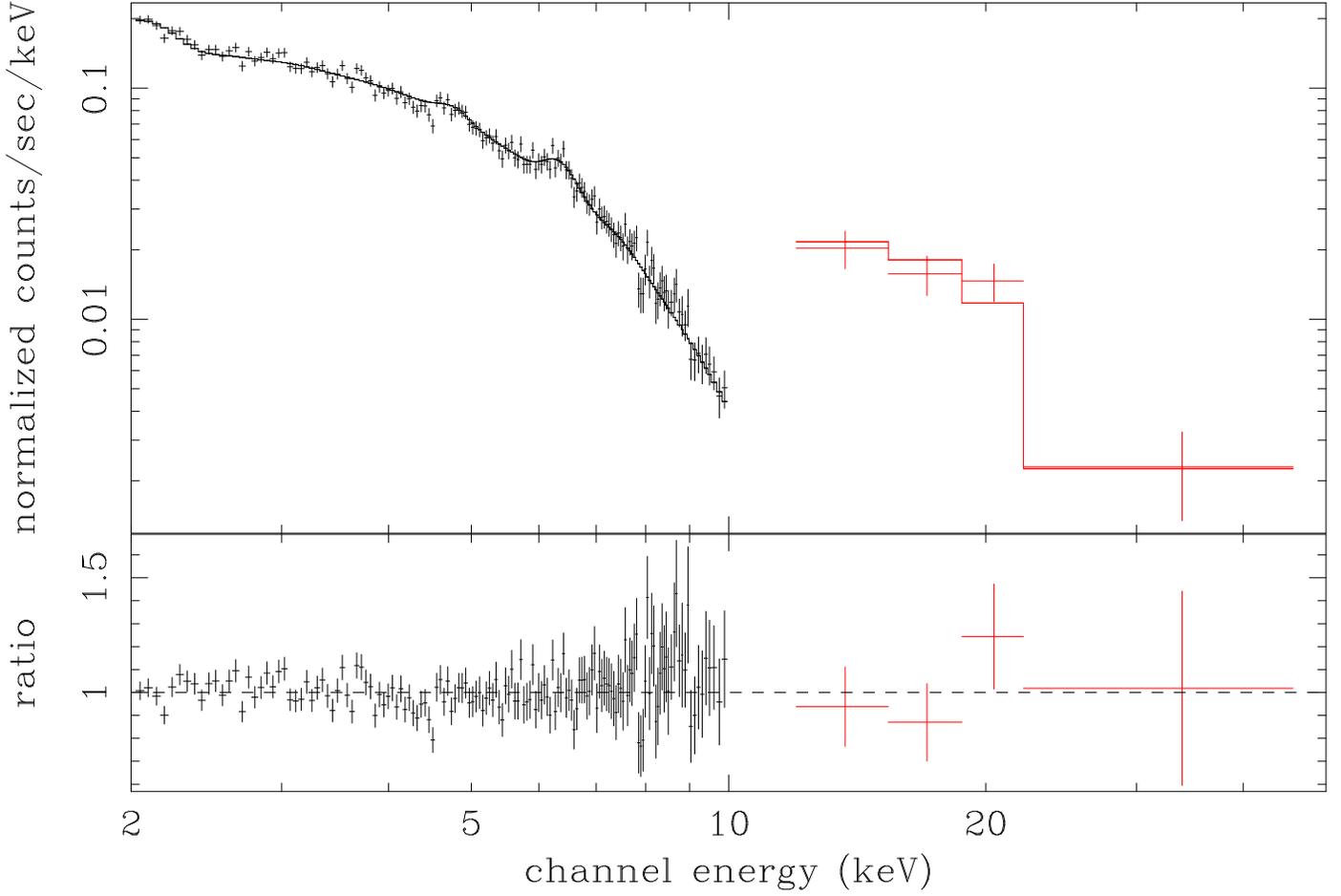}
\vskip -11cm
\figcaption {
MECS (E $<$ 10 keV) and PDS (E $>$ 10 keV) spectra and best fitting MEKAL
model for A3266.  The MECS data is extracted from a circular region
with a radius of 14$^{\prime}$ (1.2 Mpc), while
the PDS spectrum covers emission from the entire cluster.  
The difference in normalization of the
curves is due to the difference in the effective areas between the two
instruments.
}
\end{figure}}

\clearpage

\begin{figure}
\plotone{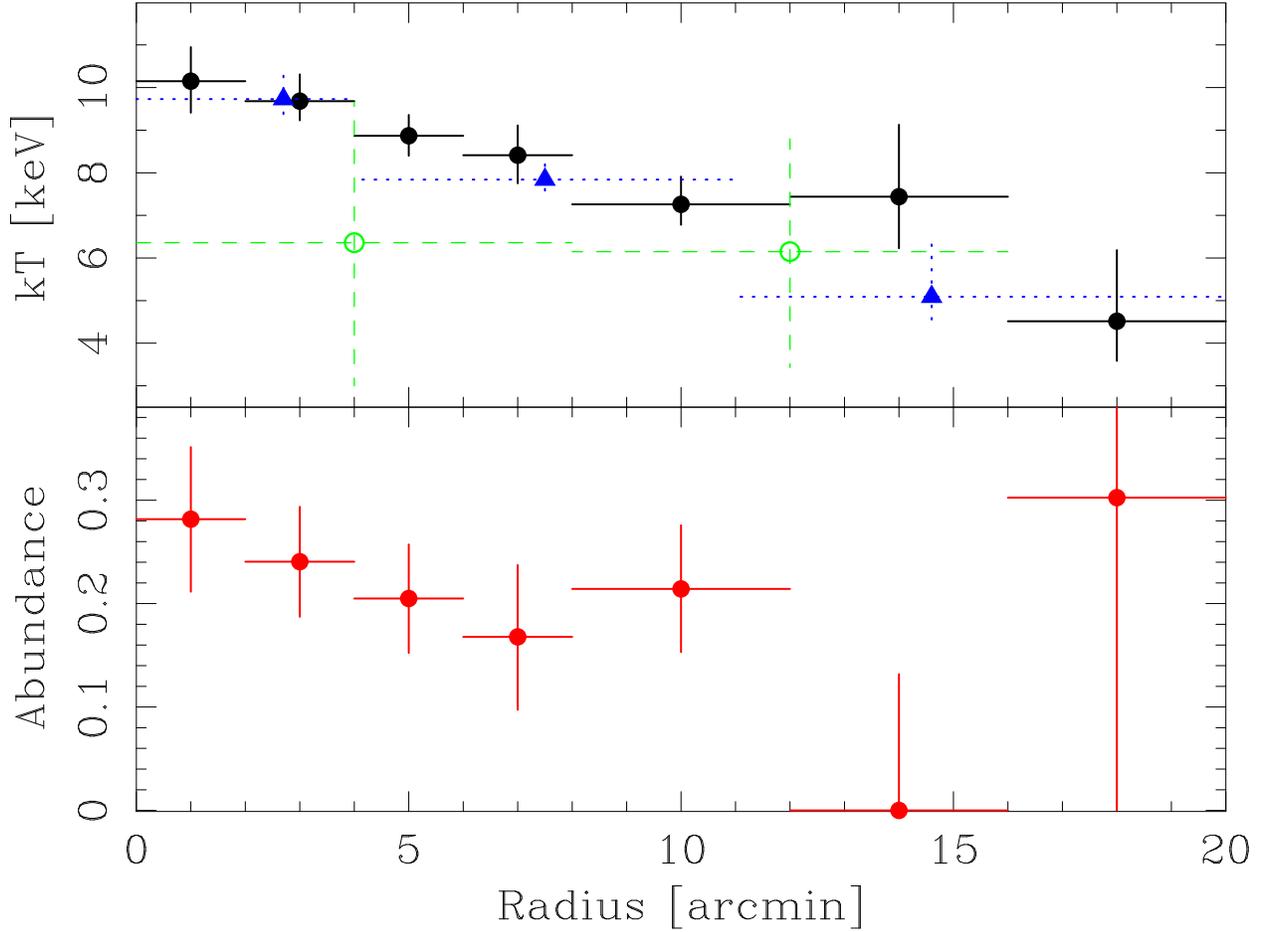}
\vskip -10cm
\caption
{{\bf Top Panel}: projected radial temperature profile from BeppoSAX
MECS data.  Filled circles indicate temperatures obtained by fitting
the continuum emission. Open circles indicate temperatures estimated
by the position of the centroid of the Fe K$_{\alpha}$ line. Filled
triangles show the temperature profile derived by Markevitch et
al. (1998), from the analysis of ASCA data. All the uncertainties on
the temperature measurements are at the 68$\%$ confidence level (we
have converted the 90$\%$ confidence errors reported in figure 4 of
Markevitch et al. 1998 into 68$\%$ confidence errors by dividing them
by 1.65, for further details see Markevitch \& Vikhlinin 1997).  
{\bf Bottom Panel}:
projected radial abundance profile from BeppoSAX MECS data. 
}
\end{figure}
\clearpage

\begin{figure}
\vskip -4cm
\plotone{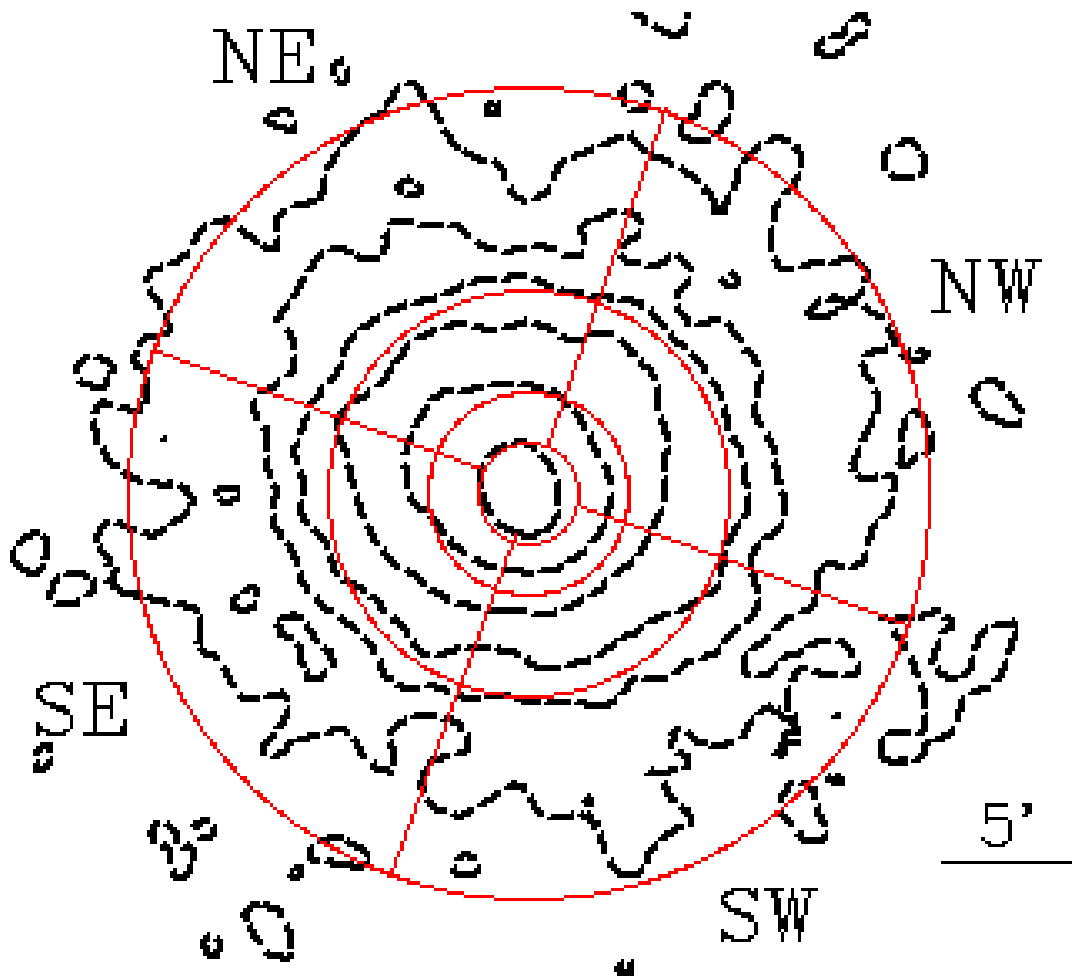}
\vskip 2cm
\caption
{BeppoSAX MECS image of A3266. Contour levels are indicated by the
dashed lines. The solid lines show how the cluster has been divided to
obtain temperature and abundance maps.}
\end{figure}
\clearpage

\begin{figure}
\vskip -1cm
\plotone{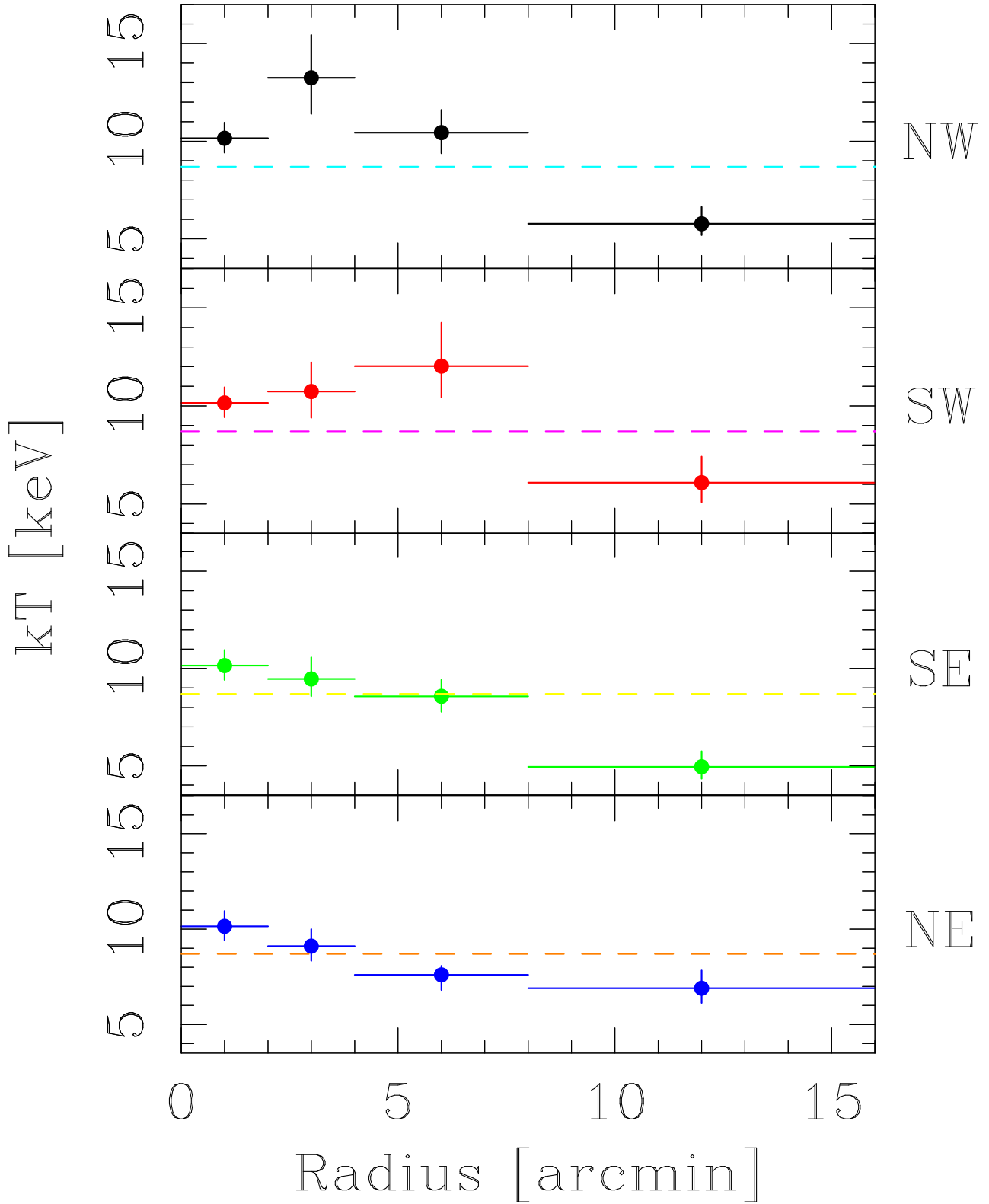}
\vskip -1cm
\caption
{Radial temperature profiles for the NW sector (first panel), the SW
sector (second panel), the SE sector (third panel) and the NE sector
(forth panel).  The temperature for the leftmost bin is derived from
the entire circle, rather than from each sector. Overlaid to each
sector as a dashed line is the cluster average temperature derived
from the radial profile shown in Fig. 2.}
\end{figure}
\clearpage


\clearpage

\end{document}